\documentclass[12pt,a4paper]{article}

  \usepackage{a4wide}
  \usepackage{latexsym}
  \usepackage{epsf}
  \usepackage{amssymb}
  \usepackage{graphicx}
  \usepackage{amsmath, cite}
  \usepackage{amsmath,amssymb,amsthm}
  \usepackage{verbatim}
  \usepackage{hyperref}
\renewcommand{\d}{\textrm{d}}

\begin{document}
\numberwithin{equation}{section}

\begin{center}
{\LARGE \bf{On classical de Sitter solutions \\
\vspace{0.3cm} in higher dimensions  }}

\vspace{1 cm} {\large Thomas Van Riet}\\

\vspace{0.7 cm}  \vspace{.15 cm} { Institut de Physique Th\'eorique,\\
CEA Saclay, CNRS URA 2306 \\ F-91191 Gif-sur-Yvette, France}

\vspace{0.7cm} {\upshape\ttfamily thomas.van-riet@cea.fr} \\

\vspace{4cm}

{\bf Abstract}
\end{center}

\begin{quotation}

We derive necessary criteria for the existence of classical
meta-stable de Sitter solutions in flux compactifications of type II
supergravity down to dimensions higher than four. We find that the
possibilities in higher dimensions are much more restricted than in
four dimensions.  The only models that satisfy the criteria are
derived from O6 compactifications to $D=5,6$ and O5
compactifications to $D=5$ and no meta-stable solutions can exist in
dimensions higher than six. All these models have in common that the
compact dimensions are negatively curved.


\end{quotation}

\newpage

\tableofcontents

\section{Introduction}

\subsection{An anti-de Sitter conspiracy theory?}
By now there is little doubt that string theory has an enormous
(possibly infinite) amount of meta-stable solutions with four large
and six small dimensions. What is less obvious is whether any of
these solutions can describe our observed universe. A popular idea
is that the high amount of solutions implies that many are close
(and one is equal) to our observed world since values for various
observable quantities are distributed around the landscape with some
non-zero probability. This vision entails a possible danger as it
ignores any subtle and non-obvious property of the landscape, if
any.

An important observable property of a solution is the size of the
cosmological constant. All \emph{explicit} and fully trustworthy
solutions that have ever been constructed in string theory have a
non-positive cosmological constant. One reason for this is that de
Sitter solutions  necessarily break all susy and are therefore more
``dirty''. However, non-susy and trustworthy vacua have been known
for some time now (see e.g.~the non-susy no-scale vacua of
\cite{Giddings:2001yu}), so the lack of susy cannot be the only
explanation. Maybe the true reason is that string theory is simply
very constraining when it comes to the existence of meta-stable de
Sitter solutions. The current state of affairs is that all de Sitter
proposals necessarily invoke quantum corrections that cannot be
computed exactly, or feature localised sources whose backreaction is
potentially problematic (see e.g.~\cite{McGuirk:2009xx,Bena:2009xk,
Douglas:2010rt, Blaback:2010sj, Bena:2010gs, Blaback:2011nz,
Giecold:2011gw, Massai:2011vi, Burgess:2011rv,Blaback:2011pn}). From
this viewpoint one can understand that all proposals for de Sitter
solutions that are claimed to be successful (which started with
\cite{Kachru:2003aw}) 
have a lack of explicitness, since the
problems hide in those places where explicit computations are
difficult. To fully appreciate this fact it is noteworthy that dS
solutions, that are not even phenomenologically interesting, have
not been constructed in a trustworthy and explicit way. Examples of
this would be de Sitter solutions in dimensions different from four
or de Sitter solutions without fine-tuned size for the cosmological
constant, all of which would be very interesting play grounds for
studying de
Sitter space in string theory nonetheless
. Hence, there seems room to believe in some ``anti-de Sitter
conspiracy'', by which we mean that the landscape of flux vacua does
not contain any meta-stable dS solution at all or, in a weaker
version, that the amount of dS vacua is dramatically less than
assumed sofar.

Disproving the strongest version of this conspiracy theory requires
only one explicit counter example that does not have to obey any
special property but being meta-stable, de Sitter and trustworthy.
It is allowed to be of any dimension, to have any susy-breaking
scale, or have only planets populated by purple aliens.

\subsection{Classical dS solutions in $D=4$?}
Perhaps the easiest place to find simple and trustworthy de Sitter
solutions is at the classical level governed by the 10-dimensional
supergravity approximation. In this setup, the Maldacena-Nunez
nogo-theorem \cite{Maldacena:2000mw} (see also \cite{deWit:1986xg})
gives us a useful lead: a solution with a time-independent, regular,
compact space without a boundary should involve brane-sources.

The idea that four-dimensional Sitter solutions could arise at the
purely classical level was revived in \cite{Hertzberg:2007wc}.
Shortly after this, some attempts were made to find explicit
solutions from orientifolds of negatively curved extra dimensions
\cite{Silverstein:2007ac, Haque:2008jz}. However these solutions can
be shown to explicitly fail to solve the 10-dimensional equations of
motion. A safer approach is to use either directly the
10-dimensional equations of motion with smeared sources, or,
equivalently, use a consistently truncated effective action in the
lower dimension. Few de Sitter solutions have been constructed
within the first \cite{Danielsson:2009ff, Danielsson:2010bc} and
second approach\cite{Flauger:2008ad, Caviezel:2008tf,
Caviezel:2009tu} (see \cite{deCarlos:2009qm, Dibitetto:2010rg} for
more related work). A review and extension of these solutions has
been presented in \cite{Danielsson:2011au}\footnote{See also
references \cite{Saltman:2004jh, Andriot:2010ju, Dong:2010pm} for
examples that break susy at the compactification scale.}. Those
solutions that were obtained with purely geometrical ingredients
(fluxes, branes and curved spaces) and for which a consistent
truncation exists that allows a trustworthy computation of moduli
masses, have all been shown to be perturbatively unstable. The known
nogo theorems against stability \cite{Covi:2008ea,
GomezReino:2008bi, Borghese:2010ei, Shiu:2011zt} are evaded in the
quoted examples and we therefore still lack a simple explanation for
the presence of tachyons. On the other hand, whenever supersymmetry
is broken there is no particular reason for a solution to be
meta-stable. Even more, if one makes the naive assumption that the
signs of moduli masses are randomly distributed then meta-stable
solutions will be rare since four-dimensional models generically
have many moduli\footnote{It is useful to consider the situation for
extended gauged supergravities. The only models that have
meta-stable de Sitter solutions occur for $\mathcal{N}\preceq 2$
supergravity \cite{Fre:2002pd, Gunaydin:2000xk, Cosemans:2005sj}
(see also \cite{Roest:2009tt}). Unfortunately the higher-dimensional
origin of these meta-stable de Sitter solutions is not known.}.

Even more, if a meta-stable solution will be found in the future,
the approximation of smeared orientifolds, might be
invalid\cite{Douglas:2010rt, Blaback:2010sj, Blaback:2011nz,
Blaback:2011pn}. In any case, at the moment there is no trustworthy,
meta-stable, classical de Sitter solution known\footnote{There are
some recent claims of simple classical warped de Sitter solutions
\cite{Neupane:2010is, Minamitsuji:2011xs}. If these examples
originate from higher dimensional supergravity they are either
necessarily singular solutions or they are to be regarded as
non-compactifications, as in \cite{Gibbons:2001wy}. It has been
explicitly shown that the de Sitter solutions of
\cite{Neupane:2010is}  are of the form of a curved brane with de
Sitter worldvolume \cite{Chemissany:2011gr}. This could still be of
relevance for a brane-world type scenario \cite{Neupane:2010ey}. }.
For now we ignore the issues of smeared orientifolds and
backreaction of localised sources and we will be satisfied with
classical de Sitter solutions with smeared sources that are
perturbatively stable.

\subsection{Classical dS solutions in $D>4$?}
Our interest for flux vacua in dimensions higher than four is mainly
coming from the increase of simplicity when one goes higher in
dimension. This simplicity arises because there are much less moduli
and the possibilities to wrap branes and fluxes in the compact
dimensions is less. Because this paper will strictly be concerned
with flux compactifications down to dimensions higher than four we
list the various motivations for this:
\begin{itemize}
\item As explained above, to disprove a hypothetical anti-de Sitter
conspiracy in string theory, it suffices to find one de Sitter
solution, which should be easier in dimensions higher than four
since there are less moduli.
\item A simple classical de Sitter solution in string theory will be
very useful as an example to study the hypothetical dS/CFT
correspondence \cite{Strominger:2001pn}. For this purpose the
dimension of the de Sitter solution does not have to be four.
\item Meta-stable de Sitter solutions in dimensions higher than four
can be dimensionally reduced to meta-stable quintessence solutions
in four dimensions \cite{Townsend:2001ea, Rosseel:2006fs}. Or in
some cases meta-stable de Sitter solutions in five dimensions can be
linked to meta-stable de Sitter solutions in four dimensions
\cite{Ogetbil:2008tk}. \item It has been shown that under certain
circumstances the decay of higher-dimensional de Sitter solutions
can give a dynamical mechanism for compactification
\cite{Carroll:2009dn}.
\end{itemize}

The rest of this paper is organised as follows. In the next section
we review some basic properties of orientifold compactifications
that will be needed. In section \ref{criteria} we derive the
necessary existence and meta-stability criteria for de Sitter
solutions in dimensions higher than four. In section \ref{dSmodels}
we discuss in some details the models that fulfill our necessary
criteria. We end with section \ref{discussion} where we discuss the
obtained results and the interesting directions for further research
they imply.

\section{Tree-level energy}

\subsection{The energy sources}
At the classical level there are three ingredients that enter the
energy of the lower-dimensional theory: fluxes, branes and curvature
of the extra dimensions. Experience with compactifications down to
four-dimensions have demonstrated that typical ingredients required
for de Sitter solutions are negative tension objects, like
orientifold planes, and negatively curved compact dimensions
\cite{Silverstein:2007ac, Haque:2008jz}. Below we will find similar
results for higher dimensions. Since orientifolds are crucial we
briefly repeat the most essential properties that we need.


An obvious restriction on the possible $Op$ planes for
compactifications down to $D$ dimensions is that $p>D-2$, otherwise
the source can not fill the $D$-dimensional space. When it comes to
models with orientifold solutions of different types one can use the
intersection rules for branes in flat space to understand what kind
of mixtures are possible\footnote{We assume that violating these
restrictions make the orientifold wrap cycles that are not
calibrated, which probably indicates an inconsistency or at least an
instability of the model.}. The following combined sources could be
possible
\begin{align}
& D=5 : O4/O8, O6/O8, O5/O9\,,\nonumber\\
& D=6 : O5/O9, O6/O8\,,\nonumber
\end{align}
and no mixture of $Op$-plane types in higher dimensions. In what
follows we discard all models with $O9$ planes since the O9 tadpole
can only be canceled by introducing sixteen D9 branes that cancel
the negative tension.

The standard way to cancel the tadpole of O$p$ planes with $p<7$,
without having to cancel the negative tension as well, uses fluxes.
The Bianchi identity for the RR $(8-p)$-form field strength, that is
magnetically sourced by an $Op$-plane, is given by
\begin{equation}
\d F_{8-p} = H\wedge F_{6-p} + J_{Op}\,,
\end{equation}
where $J_{Op}$ symbolizes the orientifold brane source and
corresponds to a singular (delta-function-like) form or to a regular
form in the smeared approximation. The tadpole condition arises when
the Bianchi identity is integrated over the compact cycle
perpendicular to the source. Clearly the integrated flux combination
$H\wedge F_{6-p}$ can cancel the charge (integrated source form) for
suitable fluxes. However, this is not possible for $O7-$ and
$O8$-planes\footnote{The same problem occurs for NS5 branes.}. The
only way to solve the tadpole condition (without loosing the
negative tension) in these cases, would be by considering pairs of
$Op$ and anti-$Op$ planes or to make the $Op$ planes wrap cycles for
which the source form $J_{Op}$ is exact\footnote{In the case of
nilmanifolds there are Op solutions known that wrap a 1-cycle whose
volume form is not closed and whose Hodge-dual form $J_{Op}$ then
becomes exact, see e.g.~\cite{Blaback:2010sj} .}.

Finally we recall the specific transformation properties of the
fluxes under the $Op$ targetspace involutions. The position of the
$Op$-plane is given by the surface that is invariant under the
target space involution $\sigma$. For all $Op$ planes the $H$-flux
must be odd $\sigma(H)= -H$. To discuss the parity of the RR form
field strengths $F_i$ we follow \cite{Koerber:2007hd} and introduce
the operator $\alpha$ that reverses the indices of a form.
Explicitly this means that
\begin{align}
&\alpha(F_i)= + F_i\,,\qquad i=0,1, 4, 5, 8, 9\,,\nonumber\\
&\alpha(F_i)= - F_i\,,\qquad i=2, 3, 6, 7\,.\nonumber
\end{align}
We then have the following parity conditions
\begin{align}
&\sigma(F_i)= +\alpha(F_i) \,,\qquad (O2, O3, O6, O7 )\,,\nonumber\\
&\sigma(F_i)= -\alpha(F_i) \,,\qquad (O0, O1, O4, O5, O8, O9 )
\,.\nonumber
\end{align}
Given that the source form $J_p$ is even for odd $p$ and odd for
even $p$ we find that the parity rules are nicely consistent with
the Bianchi-identity, which ensures that we can always employ fluxes
to cancel the charge tadpole\footnote{This corrects a wrong
statement about $O4$ models in \cite{Danielsson:2009ff}.}.

\subsection{Coupling and volume dependence of the energy}
Our starting point is the 10-dimensional supergravity action of type
II string theory in string frame
\begin{equation}
S = \int \exp(-2\phi)\sqrt{-g}\Bigl( R + (4\partial\phi)^2
-\tfrac{1}{2}\tfrac{1}{3!}H^2\Bigr) - \int
\sqrt{-g}\tfrac{1}{2}\sum_q \tfrac{1}{q!}|F_q^{RR}|^2\,,
\end{equation}
where $q$ runs over $0, 2, 4$ for type IIA sugra and over $1, 3, 5$
for type IIB supergravity, with the usual caveat for the self-dual
5-form field strength. We suppressed the Chern-Simons terms as we
will not need them. The string coupling constant is given by
$g_s=\exp(\phi)$. There are also localised sources whose action can
be added to the bulk action. The only piece we need is the DBI part
that couples to the metric. For $Op$ and $Dp$ sources this is given
by
\begin{equation}
S_{Dp/Op} = - T_p\int_p  \exp(-\phi) \sqrt{g_{p+1}}\,,
\end{equation}
where $g_{p+1}$ is the induced metric on the source worldvolume and
$T_p$ is the tension, which is negative for $Op$ sources and
positive for $Dp$ sources. We can also consider NS5 branes, whose
action differs from the $D5$ brane by an extra power of
$\exp(-\phi)$.

In the unwarped limit, the 10-dimensional metric, describing a
compactification downto $D$ dimensions, can be written as follows
\begin{equation}
\d s^2_{10} = \tau^{-2}\d s_D^2 + \rho \d s^2_{10-D}\,.
\end{equation}
The modulus $\rho$ is the string frame volume and consequently we
normalised $\int \sqrt{g_{10-D}}$ to one in string units. The
modulus $\tau$ can be shown to be
\begin{equation}
\tau^{D-2}=\exp(-2\phi)\rho^{\frac{10-D}{2}}\,,
\end{equation}
in order to have $D$-dimensional Einstein frame. The variables
$\rho$ and $\tau$ span a flat 2-dimensional subspace of the general
modulispace. In this parametrisation $\rho$ and $\tau$ do not have
standard kinetic terms, but this is not required for the analysis
done in this paper. We have chosen this specific parametrisation in
order to make contact with the previous literature on the topic.

Using the above we can derive the dependency of the various energy
contributions on the string coupling and volume modulus. The total
energy $V$ is the sum of various energies $V=\sum_i V_i$. These
separate energies $V_i$ come from the curved extra dimensions,
$V_i=V_R$, the $H$-flux, $V_i=V_H$, the RR q-form fluxes $V_i=
V_{RR}^q$, and the sources $V_i= V_{Dp/Op}$ and $V_i=V_{NS5}$.
Specifically we find
\begin{align}
& V_R \sim -\tilde{R}_{10-D}(\varphi) \rho^{-1}\tau^{-2}\,,\\
& V_H \sim |\tilde{H}|^2(\varphi) \rho^{-3}\tau^{-2}\,,\\
& V_{RR}^q \sim |\tilde{F}_q|^2(\varphi)
\rho^{\frac{10-D}{2}-q}\tau^{-D}\,,\\
& V_{Dp/Op} \sim
T_p(\varphi)\rho^{\frac{2p-D-8}{4}}\tau^{-\frac{D+2}{2}}\,,\\
&V_{NS5} \sim  T(\varphi)\rho^{-2}\tau^{-2}\,.
\end{align}
The notation is such that the tilde contractions $|\tilde{F}|^2$ are
done using the internal metric with the volume modulus $\rho$
factored out.  Furthermore we introduced the democratic notation for
fluxes that are space-filling. As an example, an $F_4$ flux filling
four non-compact dimensions (when $D=4$) will be considered through
its Hodge dual $F_6$.  We have symbolically introduced the hidden
dependence on the non-universal moduli as $\varphi$. These
non-universal moduli could be shape moduli of the internal
dimensions, or gauge potential moduli. The way they appear in the
flux contribution and source contribution depends on the details of
the cycles thread and wrapped by fluxes and sources.

In this paper we do not consider any KK monopoles or fractional
Wilson lines as in \cite{Silverstein:2007ac}. The backreaction of KK
monopoles is worry some, as well is it unclear how to find the
stable cycles for such branes. In our approach we  stick to brane
setups that lead to lower-dimensional supergravity theories. This
requires the sources to be calibrated. The benefit of this
restriction is that supersymmetry of the lower-dimensional action
restricts the possible corrections, and secondly, that the branes
are wrapped in a consistent manner.

\section{Existence and stability criteria}\label{criteria}

With the $\rho, \tau$ dependence at hand one can deduce necessary
(but not sufficient) criteria for the existence of de Sitter
critical point to the potential. This boils down to analysing the
constraints coming from the following two equations and one
inequality
\begin{equation} \label{constraints}
\partial_{\rho} V=0\,,\qquad \partial_{\tau} V=0\,, \qquad V>0\,.
\end{equation}
For the case $D=4$ this has been initiated in
\cite{Hertzberg:2007wc} and systematically worked out in
\cite{Danielsson:2009ff, Wrase:2010ew}. The equations
(\ref{constraints}) can easily shown to coincide with specific
linear combinations of the 10-dimensional dilaton equation and the
trace of the 10-dimensional Einstein equation over the compact
dimensions \cite{Danielsson:2009ff}.

The universal moduli $\rho$ and $\tau$ do not only allow us to find
existence criteria for de Sitter solutions, they also allow us to
find minimal requirements for meta-stability of the de Sitter
solutions \cite{Shiu:2011zt}. This goes as follows: if the two by
two Hessian
\begin{equation}
\begin{pmatrix}
\partial^2_{\rho} V & \partial_{\rho}\partial_{\tau} V\\
\partial_{\rho}\partial_{\tau} V & \partial^2_{\tau} V
\end{pmatrix}\,,
\end{equation}
has a negative eigenvalue it implies that the full mass matrix will
also have negative eigenvalues, through Silvesters criterium. The
eigenvalues of the above two by two Hessian will not be part of the
spectrum of the full $N$ by $N$ mass matrix due to moduli mixing,
but this is not of any importance when it comes to finding necessary
conditions for stability. In what follows it is useful to remind
that the eigenvalues $\lambda_{+}, \lambda_{-}$ of a symmetric two
by two matrix
\begin{equation}
\begin{pmatrix}
t_1 & s\\
s & t_2
\end{pmatrix}
\end{equation}
are given by
\begin{equation}
2\lambda_{\pm} = t_1 + t_2 \pm\sqrt{(t_1+t_2)^2 - 4( t_1t_2-s^2)}\,.
\end{equation}
Furthermore if either $t_1$ or $t_2$ is negative so will be at least
one of the eigenvalues due to Sylvester's criterium. The smallest
eigenvalue is always $\lambda_{-}$ and stability therefore requires
\begin{equation}\label{stability}
\lambda_{-}>0 \qquad  \Leftrightarrow \qquad \boxed{t_1t_2 > s^2}\
\,,
\end{equation}
with both $t_1$ and $t_2$ positive.  If this condition is violated
we cannot have a meta-stable de Sitter solution. Equation
(\ref{stability}) is easy to interpret. It simply states that a
negative determinant implies negative eigenvalues since the
determinant equals the product of eigenvalues.

We will now list all existence and stability criteria for de Sitter
solutions that can be obtained from the universal moduli. We
simplify the calculations using a rescaling of the moduli. Consider
a critical point of the potential at the values $\rho_c,\tau_c$ then
we redefine the variables $\rho$ and $\tau$ as follows\footnote{I am
grateful to Ulf Danielsson for pointing out this simplifying trick.}
\begin{equation}
\rho\rightarrow \frac{\rho}{\rho_c}\,,\qquad \tau\rightarrow
\frac{\tau}{\tau_c}\,.
\end{equation}
In this notation the critical point is always at the values
$\rho=1,\tau=1$. As a consequence the coefficients in the potential
are not anymore $|H|^2, |F_p|^2$ or $T$, but are rescaled by some
powers of $\rho_c$ and $\tau_c$.

\subsection{Trivial example : $D=9$ with D8/O8 sources}
Consider compactifications with only one compact direction. This
restricts the possible brane sources to be $D8/O8$ and we are
necessarily in IIA. Then the possible flux is $F_0$ flux. We write
the potential as follows
\begin{equation}
V = f_0^2\tau^{-9}\rho^{1/2} + T \tau^{-11/2}\rho^{-1/4}\,,
\end{equation}
where, up to numerical factors and rescalings with $\rho_c$ and
$\tau_c$, $f_0$ corresponds to the Romans mass and $T$ to the
tension. Then we immediately find
\begin{align}
& \partial_{\rho}V=0 \Rightarrow f_0^2 = \tfrac{1}{2}T\,,\\
& \partial_{\tau}V=0 \Rightarrow f_0^2 = -\tfrac{11}{18}T\,.
\end{align}
Hence no solution is possible at all, whether dS, AdS or Minkowski.
We have even been too mild here since, strictly speaking, the $F_0$
flux should be projected out by the $O8$ plane and we furthermore
have no way to cancel the $O8$ tadpole without canceling the $O8$
tension. In what follows we are more careful in taking into account
the orientifold involutions and tadpole conditions.

\subsection{Less trivial example: $D=7$ with D6/O6 sources}
Things start to get more interesting in $D=7$ where one possibility
is to have space-filling $D6/O6$ sources. The possible fluxes are
$F_0, F_2$ and $H$. Note that $F_2$ is odd and given that it has two
legs outside the $O6$-plane it will normally not survive the
orientifold projection. We nonetheless keep it with us, in case
there is no $O6$ plane or it can somehow survive\footnote{Note that
localised $D6/O6$ sources lead to non-trivial $F_2$ profiles, but
this is not really counted as flux, it is rather a consequence of
backreaction.}

We write the scalar potential as
\begin{equation}
V =  f_0^2 \tau^{-7}\rho^{3/2} + f_2^2\tau^{-7}\rho^{1/2} +
h^2\tau^{-2}\rho^{-3} - R\tau^{-2}\rho^{-1} + T
\tau^{-9/2}\rho^{-3/4}\,.
\end{equation}
We introduced a few new symbols: $R$ equals the curvature of the
internal dimensions (up to a positive constant), $h^2$ equals the
$H^2$ (again with some positive proportionality constant which we
will not mention anymore) and $f_2^2$ equals $F_2^2$. We then find
\begin{align}
& \partial_{\rho}V=0 \Rightarrow R = -\tfrac{3}{2}f_0^2 +\tfrac{1}{2}f_2^2 + 3 h^2 +\tfrac{3}{4}T \,,\\
& \partial_{\tau}V=0 \Rightarrow T = -\tfrac{10}{3}f_0^2 -2 f_2^2
+\tfrac{4}{3}h^2 \,.
\end{align}
If we plug this into the on-shell value for $V$ we find
\begin{equation}
V=\tfrac{5}{3}(f_0^2 -h^2)\,.
\end{equation}
This can have any sign, so this model could allow Minkowski, AdS and
dS solutions.

The Hessian is given by
\begin{equation}
\partial_i\partial_j V = \begin{pmatrix} \tfrac{35}{8}f_0^2 +\tfrac{1}{8}f_2^2
+\tfrac{23}{4}h^2 & -\tfrac{55}{4}f_0^2
-\tfrac{5}{4}f_2^2 +\tfrac{5}{2}h^2 \\
-\tfrac{55}{4}f_0^2 -\tfrac{5}{4}f_2^2 +\tfrac{5}{2}h^2 &
-\tfrac{5}{2}f_0^2 +\tfrac{25}{2}f_2^2 +15h^2
\end{pmatrix}\,.
\end{equation}
When $V=0$ (and $f_2^2=0$) we find that $R=0$ and that the Hessian
has one positive and zero eigenvalue, consistent with the fact that
the scalar potential can then be written as a square
\begin{equation}
V = f_0^2(\tau^{-7/2}\rho^{3/4} - \tau^{-1}\rho^{-3/2} )^2\,.
\end{equation}
These solutions are the no-scale Minkowski solutions constructed in
\cite{Blaback:2010sj} and they exist in dimensions $D=2\,\ldots 7$
and were first studied in $D=4$ \cite{Giddings:2001yu}.

In order to have de Sitter solutions we need $f_0^2 > h^2$ and in
that case we can demonstrate that the tension is necessarily
negative, which corresponds to having net $O6$ sources. Let us
therefore take $f_2^2=0$ in what follows. The determinant of the
Hessian is given by
\begin{equation}
det(\partial_i\partial_j V) = 40f_0^4(-5  + 2\frac{h^4}{f_0^4} + 3
\frac{h^2}{f_0^2})\,.
\end{equation}
De Sitter requires $h^2/f_0^2<1$ with the no-scale Minkowski
solution at the turning point $h^2/f_0^2=1$.  Hence we find an
elegant structure: \emph{exactly at the Minkowski turning point, a
tachyon appears and the de Sitter critical points can never be a
local minima of the potential.} These unstable de Sitter solutions
could possible be engineered with pairs of $O6$ and anti-$O6$ planes
as pointed out in \cite{Blaback:2011nz}.

\subsection{Summary of all possibilities}
As shown in the two examples the technique is to use the two
$\partial V=0$ equations to eliminate $R$ and $T$ in terms of the
fluxes, which are strict positive. Then this is plugged into the
on-shell value for $V$ to read of the sign of the cosmological
constant. To determine stability we do the same for the Hessian.

Let us summarise the result of the computation:
\begin{enumerate}
\item $D>7$: No dS critical points are possible. \item $D=7$:
A dS critical point build from $O6$ planes and negative curvature is
allowed but necessarily unstable as shown explicitly in the previous
section.
\item $D=6$: $O5$ sources with negatively curved compact dimensions allow
dS critical points, which are necessarily unstable. However $O6$
sources can have critical points which are perturbatively stable in
the $\rho,\tau$ directions if $F_2$ is large enough. The curvature
of the internal dimensions is required to be negative.
\item $D=5$: $O4$ models with negatively curved compact dimensions
again only allow unstable dS critical points, whereas $O5$ and $O6$
models can evade unstable directions at a dS critical point. Also
here the curvature of the internal dimensions is negative.
\end{enumerate}
Most relevant to observe here is that meta-stable solutions in $D>6$
cannot exist and that for $D=5, 6$ there is only a  small amount of
models that potentially allow meta-stable de Sitter solutions.

There is a recurring pattern in each dimension $D$ that allows a
no-scale Minkowski solution with space-filling $O(D-1)$-planes
(which are the values $D=2,\ldots, 7$ \cite{Blaback:2010sj}). These
models all allow de Sitter critical points by negatively curving the
compact dimensions of the no-scale solution and changing the ratio
between net tension and charge of the orientifolds (such that there
is more net negative tension than net negative charge). These
solutions always have an unstable direction, which coincides with
the massless direction of the no-scale Minkowski solution. In other
words, the results of section 3.2 for the case of $O6$ planes in
$D=7$ extends to the other dimensions as well.

We have furthermore checked that none of the mixed orientifold plane
combinations mentioned in section 2 fulfill the criteria. This is
straightforward for the $O5/09$ combination since we discarded
solutions with net $O9$ tension as the $O9$ tadpole cannot be
canceled without canceling the $O9$ tension. The $O6/O8$
combinations naively fulfill the criteria in $D=5$ and $D=6$ since
$O6$ planes separately do. However the presence of the $O8$ plane
implies that the necessary $F_0$ flux is projected out and that
tadpole constraint cannot be solved easily. The same tadpole
constraint problem is there for the $O4/O8$ model in $D=5$. Upon
neglecting tadpole constraints the $O4/O8$ combination does satisfy
the criteria for de Sitters solutions that are stable in the
$(\rho,\tau)$-directions.

\section{dS building grounds}\label{dSmodels}

In this section we provide some details of the models in $D=5$ and
$D=6$ that potentially allow meta-stable dS solutions.

\subsection{dS model I: $O6$ planes in $D=6$}

In this case the $O6$ planes wrap a one-cycle which projects out any
$F_4$ flux, since the flux would have one leg along the 1-cycle
wrapped by the $O6$, giving it odd parity. The remaining fluxes are
$H$, $F_0$ and $F_2$. The $\partial V =0$ equations lead to
\begin{align}
& R = \frac{11}{3}(h^2 - f_0^2) - f^2_2\,,\\
& 3T = -10 f_0^2 -6f_2^2 + 4 h^2\,.
\end{align}
When plugged into the potential we find
\begin{equation}
V = \frac{4}{3}(f_0^2 - h^2)\,,
\end{equation}
which shows that dS solutions require $f^2_0>h^2$ and imply negative
internal curvature $R<0$ and net $O6$-plane tension $T<0$.

The necessary stability condition (\ref{stability}) requires us to
compute the determinant of the Hessian. The Hessian is given by
\begin{equation}
\partial_i\partial_j V =\begin{pmatrix}\frac{41}{6}f_0^2 + \frac{17}{3}h^2 + \frac{1}{2}f_2^2 & -\frac{34}{3}f_0^2 + \frac{4}{3}h^2 - 2f_2^2 \\
-\frac{34}{3}f_0^2 + \frac{4}{3}h^2 - 2f_2^2 & -\frac{8}{3}f_0^2 +
\frac{32}{3}h^2 + 8 f_2^2
\end{pmatrix}\,.
\end{equation}
For the Minkowski solutions $f_0^2 = h^2$, the determinant of the
Hessian simplifies to
\begin{equation}
det(\partial_i\partial_j V) = 64 f_0^2f_2^2\,,
\end{equation}
which demonstrates that the Minkowski solutions are stable in the
$\rho,\tau$-directions, and that the solution is no-scale when
either $f_0$ or $f_2$ vanish. For the case $f_0^2 = h^2 =0$ and
$f_2\neq 0$ this no-scale solution has been explicitly constructed
in \cite{Blaback:2010sj} from  T-dualising GKP\cite{Giddings:2001yu}
and this solution  corresponds to an $O6$ wrapping a one-cycle in a
nilmanifold.

We can verify that the Hessian can be positive definite for de
Sitter solutions by slightly perturbing the Minkowski solutions to:
\begin{equation}
f_0^2 = h^2  +\delta\,,\qquad \delta>0\,.
\end{equation}
Where we think of $\delta$ as arbitrary tiny and positive. At
first-order in $\delta$ the determinant then becomes
\begin{equation}
det(\partial_i\partial_j V) = 64 h^2f_2^2 -\frac{616}{3}h^2\delta +
8f_2^2\delta \,.
\end{equation}
Hence it is a simple consequence of continuity of the determinant of
the Hessian that de Sitter solutions that are stable in the
($\rho,\tau$)-directions exist when perturbing the no-scale
Minkowski vacua given by $f_0=h=0$ and $f_2\neq 0$. This is an
interesting place to look for dS solutions since solutions could be
tuned very close to a (susy) Minkowski solution.

\subsection{dS model II: $O5$ planes in $D=5$}
This situation is as good as identical to the O6 planes in $D=6$,
with the roles of the $F_0$ and $F_2$ flux now played by the $F_1$
and the $F_3$ fluxes. First one observes that the $F_5$-flux must be
projected out (just like the $F_4$ before). The $\partial V=0$
equations lead to
\begin{align}
&R = \frac{9}{2}h^2 - \frac{9}{2}f_1^2 - f_3^2\,,\\
&T = 2h^2 - 4f_1^2 - 2 f_3^2\,,\\
&V = \frac{3}{2}(f_1^2 - h^2)\,.
\end{align}
Again this implies that de Sitter solutions have negatively curved
internal dimensions and net orientifold tension.  Minkowski
solutions exist whenever $f_1 = h$. The Hessian is given by
\begin{equation}
\partial_i\partial_j V =\begin{pmatrix}\frac{45}{8}h^2 +\frac{9}{2}f_1^2 +\tfrac{1}{8}f_3^2  & \frac{9}{4}h^2 - 9 f_1^2 -\frac{3}{4}f_3^2 \\
\frac{9}{4}h^2 - 9 f_1^2 -\frac{3}{4}f_3^2 & \frac{21}{2}h^2 - 6
f_1^2 +\frac{9}{2}f_3^2
\end{pmatrix}\,.
\end{equation}
This Hessian has the same structure as the example above. When
slightly perturbing the Minkowski solutions towards de Sitter
solutions by $f_1^2 = h^2 +\delta$, where $\delta>0$, the Hessian
simplifies and the determinant becomes
\begin{equation}
det(\partial_i\partial_j V) = 36 h^2f_3^2 - 162h^2\delta +
6f_3^2\delta\,.
\end{equation}
We notice that the perturbation of the no-scale solutions with with
$f_1=h=0$ (and $f_3\neq 0$) can give de Sitter solutions which are
stable in the $(\rho,\tau)$-directions.

\subsection{dS model III: $O6$ planes in $D=5$}
The two $\partial V =0$ equations entail
\begin{align}
& R =  \frac{10}{3}h^2 -\frac{10}{3}f_0^2 - f_2^2 +\frac{4}{3}f_4^2\,,\\
& T = \frac{4}{3}h^2 -\frac{10}{3}f_0^2 -2 f_2^2
-\frac{2}{3}f_4^2\,,\\
&V = f_0^2 - h^2 - f_4^2\,.
\end{align}
As before, we find that de Sitter solutions necessarily require
negative curvature ($R<0$) and  negative tension ($T<0$). The
Hessian is given by
\begin{equation}
\partial_i\partial_j V = \begin{pmatrix} \frac{23}{4}h^2 + \frac{75}{8}f_0^2 +\frac{9}{8}f_2^2 + \frac{7}{8}f_4^2  & \frac{1}{2}h^2 -\frac{35}{4}f_0^2 -\frac{9}{4}f_2^2 +\frac{17}{4}f_4^2\\
\frac{1}{2}h^2 -\frac{35}{4}f_0^2 -\frac{9}{4}f_2^2
+\frac{17}{4}f_4^2 & 7h^2 -\frac{5}{2}f^2_0 +\frac{9}{2}f_2^2
+\frac{23}{2}f_4^2\end{pmatrix}\,.
\end{equation}
The Minkowski solutions are defined by
\begin{equation}
f_0^2 = h^2 + f_4^2\,.
\end{equation}
In this case the determinant of the Hessian is given by
\begin{equation}
det(\partial_i\partial_j V) = 9 ( 8f_4^4 + 4h^2f_2^2 + 12 h^2f_4^2 +
4 f_4^2f_2^2) \,.
\end{equation}
Hence dS points close by are stable in the $(\rho,
\tau)$-directions. Note that no-scale Minkowski solutions are
defined by $h^2 = f_0^2 = f_4^2$ and only $f_2^2\neq 0$. Perturbing
those no-scale solutions into the dS regime by $f_0^2 =\delta$ leads
to an instability in the $(\rho, \tau)$-directions.

\section{Discussion}\label{discussion}

\subsection{Obtained results}
In this paper we derived necessary (but not sufficient) conditions
for the existence of classical meta-stable de Sitter solutions in
$D>4$ dimensions from orientifold compactifications of type II
supergravity. The necessary conditions are derived from the
universal dependence of the scalar potential on the dilaton and
volume modulus. This is a continuation of the results known for
compactifications to $D=4$, as derived in \cite{Hertzberg:2007wc,
Danielsson:2009ff, Wrase:2010ew, Shiu:2011zt}, for which we have
simplified the method significantly.

The results show that no meta-stable de Sitter solution in $D>6$ can
exist and that only few possibilities in $D=5, 6$ can work. The
three cases that fulfill the necessary criteria ($O6$ in $D=5, 6$
and $O5$ in $D=5$) have in common that the net source tension has to
be negative and that the curvature of the internal dimensions has to
be negative. This is in line with almost all examples in
$D=4$\footnote{We have noticed  that the $D=4$ examples were
negative curvature is not required probably require fluxes that are
projected out by the orientifold or require difficult-to-satisfy
tadpole constraints.}.

Some of the criteria can be weakened when wrapped NS5 branes are
included. We have not explicitly listed the new possibilities that
arise with NS5 branes, because NS5 branes are difficult to
incorporate in simple models. This comes from the NS5 tadpole
condition, which cannot be satisfied using fluxes. Therefore one
necessarily has to wrap the NS5 branes on trivial cycles or allow
anti-NS5 branes in homologous cycles without the branes
annihilating. Such possibilities are not easy to construct in a
trustworthy manner \cite{Conlon:2011qp}.

An important drawback of our derivation is the smeared (unwarped)
approximation. This approximation works for deriving  BPS solutions
\cite{Giddings:2001yu, Blaback:2010sj}, but could easily be
problematic for non-BPS solutions \cite{Blaback:2011nz,
Blaback:2011pn}. This is especially easy to spot from the
10-dimensional Einstein equations for dS solutions from orientifolds
of negatively curved compact dimensions \cite{Douglas:2010rt}, which
comprises all our examples.

\subsection{Interesting problems for future research}
The obtained results suggest many different directions for further
research, which we list here:
\begin{itemize}
\item  Both the $O6$
in $D=6$ model and the $O5$ in $D=5$ model have Minkowski vacua and
the would-be de Sitter solutions close to those Minkowski vacua are
stable in the coupling and volume directions.  However, preliminary
investigations similar to the ones in \cite{Blaback:2010sj} indicate
that dS solutions might be excluded in these models\footnote{This is
similar to the way dS solutions can be excluded in IIB supergravity
with BPS $D3/O3$ sources \cite{Giddings:2001yu}}. This would point
to the very exciting possibility that only the $O6$ model in $D=5$
is left.

\item Our criteria are derived from the volume modulus and dilaton.
But clearly, for every scalar field that is added one obtains a new
stability and existence criterium. It should be possible to add one
more ``universal'' scalar field, without having to fix the geometry
and topology of the model. Such a scalar field could be the volume
of the cycle wrapped by the orientifold.

\item An obvious method that presents itself is to classify
4 and 5 dimensional group manifolds that allow the proper $O5$ and
$O6$ involutions, similar to the investigation in
\cite{Danielsson:2011au, Andriot:2010ju}. This allows to scan a
large set of models for de Sitter solutions. We hope to report on
this in the future.

\item It would be useful to compare the existence and stability issues for de
Sitter solutions from type II orientifolds with those from
$\alpha'$-corrections in heterotic supergravity as these should be
dual to each other\footnote{I would like to thank Callum Quigley for
pointing this out.}. In this respect the results of the recent paper
\cite{Green:2011cn} indicate that meta-stable de Sitter solutions
can be ruled out in large regions of parameter space.

\end{itemize}

\section*{Acknowledgments}
I have benefited from discussions with Ulf Danielsson and Timm Wrase
and I am especially grateful to Timm Wrase for spotting errors and
typos in the first draft. My work is supported by the ERC Starting
Independent Researcher Grant 259133-ObservableString.

\bibliography{refs}

\providecommand{\href}[2]{#2}\begingroup\raggedright\begin{thebibliography}{10}

\bibitem{Giddings:2001yu}
S.~B. Giddings, S.~Kachru and J.~Polchinski,  {\em {Hierarchies from fluxes in
  string compactifications}}, Phys. Rev. {\bf D66} (2002) 106006
[\href{http://www.arXiv.org/abs/hep-th/0105097}{{\tt hep-th/0105097}}].

\bibitem{McGuirk:2009xx}
P.~McGuirk, G.~Shiu and Y.~Sumitomo,  {\em {Non-supersymmetric infrared
  perturbations to the warped deformed conifold}}, Nucl.Phys. {\bf B842} (2010)
  383--413 [\href{http://www.arXiv.org/abs/0910.4581}{{\tt 0910.4581}}].

\bibitem{Bena:2009xk}
I.~Bena, M.~Grana and N.~Halmagyi,  {\em {On the Existence of Meta-stable Vacua
  in Klebanov-Strassler}}, JHEP {\bf 1009} (2010) 087
  [\href{http://www.arXiv.org/abs/0912.3519}{{\tt 0912.3519}}].

\bibitem{Douglas:2010rt}
M.~R. Douglas and R.~Kallosh,  {\em {Compactification on negatively curved
  manifolds}}, JHEP {\bf 06} (2010) 004
[\href{http://www.arXiv.org/abs/1001.4008}{{\tt 1001.4008}}].

\bibitem{Blaback:2010sj}
J.~Blaback, U.~H.~Danielsson, D.~Junghans, T.~Van~Riet, T.~Wrase and
  M.~Zagermann,  {\em {Smeared versus localised sources in flux
  compactifications}}, JHEP {\bf 12} (2010) 043
[\href{http://www.arXiv.org/abs/1009.1877}{{\tt 1009.1877}}].

\bibitem{Bena:2010gs}
I.~Bena, G.~Giecold and N.~Halmagyi,  {\em {The Backreaction of Anti-M2 Branes
  on a Warped Stenzel Space}}, JHEP {\bf 04} (2011) 120
[\href{http://www.arXiv.org/abs/1011.2195}{{\tt 1011.2195}}].

\bibitem{Blaback:2011nz}
J.~Blaback, U.~H.~Danielsson, D.~Junghans, T.~Van~Riet, T.~Wrase and
  M.~Zagermann,  {\em {The problematic backreaction of SUSY-breaking branes}},
\href{http://www.arXiv.org/abs/1105.4879}{{\tt 1105.4879}}.

\bibitem{Giecold:2011gw}
G.~Giecold, E.~Goi and F.~Orsi,  {\em {Assessing a candidate IIA dual to
  metastable supersymmetry-breaking}},
\href{http://www.arXiv.org/abs/1108.1789}{{\tt 1108.1789}}.

\bibitem{Massai:2011vi}
S.~Massai,  {\em {Metastable Vacua and the Backreacted Stenzel Geometry}},
\href{http://www.arXiv.org/abs/1110.2513}{{\tt 1110.2513}}.

\bibitem{Burgess:2011rv}
C.~P. Burgess, A.~Maharana, L.~van Nierop, A.~A. Nizami and F.~Quevedo,  {\em
  {On Brane Back-Reaction and de Sitter Solutions in Higher- Dimensional
  Supergravity}},
\href{http://www.arXiv.org/abs/1109.0532}{{\tt 1109.0532}}.

\bibitem{Blaback:2011pn}
J.~Blaback {\em et al.},  {\em {(Anti-)Brane backreaction beyond perturbation
  theory}},
\href{http://www.arXiv.org/abs/1111.2605}{{\tt 1111.2605}}.

\bibitem{Kachru:2003aw}
S.~Kachru, R.~Kallosh, A.~D. Linde and S.~P. Trivedi,  {\em {De Sitter vacua in
  string theory}}, Phys.Rev. {\bf D68} (2003) 046005
  [\href{http://www.arXiv.org/abs/hep-th/0301240}{{\tt hep-th/0301240}}].

\bibitem{Maldacena:2000mw}
J.~M. Maldacena and C.~Nunez,  {\em {Supergravity description of field theories
  on curved manifolds and a no go theorem}}, Int. J. Mod. Phys. {\bf A16}
  (2001) 822--855
[\href{http://www.arXiv.org/abs/hep-th/0007018}{{\tt hep-th/0007018}}].

\bibitem{deWit:1986xg}
B.~de~Wit, D.~J. Smit and N.~D. Hari~Dass,  {\em {Residual Supersymmetry of
  Compactified D=10 Supergravity}}, Nucl. Phys. {\bf B283} (1987)
165.

\bibitem{Hertzberg:2007wc}
M.~P. Hertzberg, S.~Kachru, W.~Taylor and M.~Tegmark,  {\em {Inflationary
  constraints on type IIA string theory}}, JHEP {\bf 12} (2007) 095
[\href{http://www.arXiv.org/abs/0711.2512}{{\tt 0711.2512}}].

\bibitem{Silverstein:2007ac}
E.~Silverstein,  {\em {Simple de Sitter Solutions}}, Phys. Rev. {\bf D77}
  (2008) 106006
[\href{http://www.arXiv.org/abs/0712.1196}{{\tt 0712.1196}}].

\bibitem{Haque:2008jz}
S.~S. Haque, G.~Shiu, B.~Underwood and T.~Van~Riet,  {\em {Minimal simple de
  Sitter solutions}}, Phys. Rev. {\bf D79} (2009) 086005
[\href{http://www.arXiv.org/abs/0810.5328}{{\tt 0810.5328}}].

\bibitem{Danielsson:2009ff}
U.~H. Danielsson, S.~S. Haque, G.~Shiu and T.~Van~Riet,  {\em {Towards
  classical de Sitter solutions in string theory}}, JHEP {\bf 09} (2009) 114
[\href{http://www.arXiv.org/abs/0907.2041}{{\tt 0907.2041}}].

\bibitem{Danielsson:2010bc}
U.~H. Danielsson, P.~Koerber and T.~Van~Riet,  {\em {Universal de Sitter
  solutions at tree-level}}, JHEP {\bf 05} (2010) 090
[\href{http://www.arXiv.org/abs/1003.3590}{{\tt 1003.3590}}].

\bibitem{Flauger:2008ad}
R.~Flauger, S.~Paban, D.~Robbins and T.~Wrase,  {\em {Searching for slow-roll
  moduli inflation in massive type IIA supergravity with metric fluxes}}, Phys.
  Rev. {\bf D79} (2009) 086011
[\href{http://www.arXiv.org/abs/0812.3886}{{\tt 0812.3886}}].

\bibitem{Caviezel:2008tf}
C.~Caviezel, P.~Koerber, S.~K{\"o}rs, D.~L{\"u}st, T.~Wrase and M.~Zagermann,
  {\em {On the cosmology of type IIA compactifications on SU(3)-structure
  Manifolds}}, JHEP {\bf 04} (2009) 010
[\href{http://www.arXiv.org/abs/0812.3551}{{\tt 0812.3551}}].

\bibitem{Caviezel:2009tu}
C.~Caviezel, T.~Wrase and M.~Zagermann,  {\em {Moduli stabilization and
  cosmology of type IIB on SU(2)-structure orientifolds}}, JHEP {\bf 04} (2010)
  011
[\href{http://www.arXiv.org/abs/0912.3287}{{\tt 0912.3287}}].

\bibitem{deCarlos:2009qm}
B.~de~Carlos, A.~Guarino and J.~M. Moreno,  {\em {Complete classification of
  Minkowski vacua in generalised flux models}}, JHEP {\bf 02} (2010) 076
[\href{http://www.arXiv.org/abs/0911.2876}{{\tt 0911.2876}}].

\bibitem{Dibitetto:2010rg}
G.~Dibitetto, R.~Linares and D.~Roest,  {\em {Flux compactifications, gauge
  algebras and de Sitter}}, Phys. Lett. {\bf B688} (2010) 96--100
[\href{http://www.arXiv.org/abs/1001.3982}{{\tt 1001.3982}}].

\bibitem{Danielsson:2011au}
U.~H. Danielsson, S.~S. Haque, P.~Koerber, G.~Shiu, T.~Van~Riet {\em et al.},
  {\em {De Sitter hunting in a classical landscape}}, Fortsch.Phys. {\bf 59}
  (2011) 897--933 [\href{http://www.arXiv.org/abs/1103.4858}{{\tt 1103.4858}}].

\bibitem{Saltman:2004jh}
A.~Saltman and E.~Silverstein,  {\em {A new handle on de Sitter
  compactifications}}, JHEP {\bf 01} (2006) 139
[\href{http://www.arXiv.org/abs/hep-th/0411271}{{\tt hep-th/0411271}}].

\bibitem{Andriot:2010ju}
D.~Andriot, E.~Goi, R.~Minasian and M.~Petrini,  {\em {Supersymmetry breaking
  branes on solvmanifolds and de Sitter vacua in string theory}},
\href{http://www.arXiv.org/abs/1003.3774}{{\tt 1003.3774}}.

\bibitem{Dong:2010pm}
X.~Dong, B.~Horn, E.~Silverstein and G.~Torroba,  {\em {Micromanaging de Sitter
  holography}}, Class. Quant. Grav. {\bf 27} (2010) 245020
[\href{http://www.arXiv.org/abs/1005.5403}{{\tt 1005.5403}}].

\bibitem{Covi:2008ea}
L.~Covi {\em et al.},  {\em {de Sitter vacua in no-scale supergravities and
  Calabi-Yau string models}}, JHEP {\bf 06} (2008) 057
[\href{http://www.arXiv.org/abs/0804.1073}{{\tt 0804.1073}}].

\bibitem{GomezReino:2008bi}
M.~Gomez-Reino, J.~Louis and C.~A. Scrucca,  {\em {No metastable de Sitter
  vacua in N=2 supergravity with only hypermultiplets}}, JHEP {\bf 02} (2009)
  003
[\href{http://www.arXiv.org/abs/0812.0884}{{\tt 0812.0884}}].

\bibitem{Borghese:2010ei}
A.~Borghese and D.~Roest,  {\em {Metastable supersymmetry breaking in extended
  supergravity}}, JHEP {\bf 05} (2011) 102
[\href{http://www.arXiv.org/abs/1012.3736}{{\tt 1012.3736}}].

\bibitem{Shiu:2011zt}
G.~Shiu and Y.~Sumitomo,  {\em {Stability Constraints on Classical de Sitter
  Vacua}}, \href{http://www.arXiv.org/abs/1107.2925}{{\tt 1107.2925}}.

\bibitem{Fre:2002pd}
P.~Fre, M.~Trigiante and A.~Van~Proeyen,  {\em {Stable de Sitter vacua from N =
  2 supergravity}}, Class. Quant. Grav. {\bf 19} (2002) 4167--4194
[\href{http://www.arXiv.org/abs/hep-th/0205119}{{\tt hep-th/0205119}}].

\bibitem{Gunaydin:2000xk}
M.~Gunaydin and M.~Zagermann,  {\em {The Vacua of 5-D, N=2 gauged
  Yang-Mills/Einstein tensor supergravity: Abelian case}}, Phys.Rev. {\bf D62}
  (2000) 044028 [\href{http://www.arXiv.org/abs/hep-th/0002228}{{\tt
  hep-th/0002228}}].

\bibitem{Cosemans:2005sj}
B.~Cosemans and G.~Smet,  {\em {Stable de Sitter vacua in N = 2, D = 5
  supergravity}}, Class.Quant.Grav. {\bf 22} (2005) 2359--2380
  [\href{http://www.arXiv.org/abs/hep-th/0502202}{{\tt hep-th/0502202}}].

\bibitem{Roest:2009tt}
D.~Roest and J.~Rosseel,  {\em {De Sitter in Extended Supergravity}}, Phys.
  Lett. {\bf B685} (2010) 201--207
[\href{http://www.arXiv.org/abs/0912.4440}{{\tt 0912.4440}}].

\bibitem{Neupane:2010is}
I.~P. Neupane,  {\em {Warped compactification on curved manifolds}}, Class.
  Quant. Grav. {\bf 28} (2011) 125015
[\href{http://www.arXiv.org/abs/1006.4495}{{\tt 1006.4495}}].

\bibitem{Minamitsuji:2011xs}
M.~Minamitsuji and K.~Uzawa,  {\em {Warped de Sitter compactifications and
  modulus stabilization}},
\href{http://www.arXiv.org/abs/1109.4818}{{\tt 1109.4818}}.

\bibitem{Gibbons:2001wy}
G.~W. Gibbons and C.~M. Hull,  {\em {de Sitter space from warped supergravity
  solutions}},
\href{http://www.arXiv.org/abs/hep-th/0111072}{{\tt hep-th/0111072}}.

\bibitem{Chemissany:2011gr}
W.~Chemissany, B.~Janssen and T.~Van~Riet,  {\em {Einstein Branes}}, JHEP {\bf
  10} (2011) 002
[\href{http://www.arXiv.org/abs/1107.1427}{{\tt 1107.1427}}].

\bibitem{Neupane:2010ey}
I.~P. Neupane,  {\em {De Sitter brane-world, localization of gravity, and the
  cosmological constant}}, Phys. Rev. {\bf D83} (2011) 086004
[\href{http://www.arXiv.org/abs/1011.6357}{{\tt 1011.6357}}].

\bibitem{Strominger:2001pn}
A.~Strominger,  {\em {The dS/CFT correspondence}}, JHEP {\bf 10} (2001) 034
[\href{http://www.arXiv.org/abs/hep-th/0106113}{{\tt hep-th/0106113}}].

\bibitem{Townsend:2001ea}
P.~K. Townsend,  {\em {Quintessence from M-theory}}, JHEP {\bf 11} (2001) 042
[\href{http://www.arXiv.org/abs/hep-th/0110072}{{\tt hep-th/0110072}}].

\bibitem{Rosseel:2006fs}
J.~Rosseel, T.~Van~Riet and D.~B. Westra,  {\em {Scaling cosmologies of N = 8
  gauged supergravity}}, Class. Quant. Grav. {\bf 24} (2007) 2139--2152
[\href{http://www.arXiv.org/abs/hep-th/0610143}{{\tt hep-th/0610143}}].

\bibitem{Ogetbil:2008tk}
O.~Ogetbil,  {\em {Stable de Sitter Vacua in 4 Dimensional Supergravity
  Originating from 5 Dimensions}}, Phys. Rev. {\bf D78} (2008) 105001
[\href{http://www.arXiv.org/abs/0809.0544}{{\tt 0809.0544}}].

\bibitem{Carroll:2009dn}
S.~M. Carroll, M.~C. Johnson and L.~Randall,  {\em {Dynamical compactification
  from de Sitter space}}, JHEP {\bf 11} (2009) 094
[\href{http://www.arXiv.org/abs/0904.3115}{{\tt 0904.3115}}].

\bibitem{Koerber:2007hd}
P.~Koerber and D.~Tsimpis,  {\em {Supersymmetric sources, integrability and
  generalized-structure compactifications}}, JHEP {\bf 08} (2007) 082
[\href{http://www.arXiv.org/abs/0706.1244}{{\tt 0706.1244}}].

\bibitem{Wrase:2010ew}
T.~Wrase and M.~Zagermann,  {\em {On classical de Sitter vacua in string
  theory}}, Fortschr. Phys. {\bf 58} (2010) 906--910
[\href{http://www.arXiv.org/abs/1003.0029}{{\tt 1003.0029}}].

\bibitem{Conlon:2011qp}
J.~P. Conlon,  {\em {Brane-Antibrane Backreaction in Axion Monodromy
  Inflation}},
\href{http://www.arXiv.org/abs/1110.6454}{{\tt 1110.6454}}.

\bibitem{Green:2011cn}
S.~R. Green, E.~J. Martinec, C.~Quigley and S.~Sethi,  {\em {Constraints on
  String Cosmology}}, \href{http://www.arXiv.org/abs/1110.0545}{{\tt
  1110.0545}}, * Temporary entry *.

\end{thebibliography}\endgroup

\bibliographystyle{utphysmodb}
\end{document}